\documentclass[iop,revtex4]{emulateapj}
\usepackage{graphicx}
\usepackage{natbib}
\usepackage{adjustbox}
\usepackage{rotating}

\shorttitle{$z \ge$~1 3CRRs radio galaxies vs. quasars}
\shortauthors{Marin \& Antonucci}

\begin{document}
\title{A robust derivation of the tight relationship of radio core dominance to inclination angle in high redshift 3CRR sources}

\author{Fr\'ed\'eric Marin}
\affil{Observatoire Astronomique de Strasbourg, Universit\'e de Strasbourg, CNRS, UMR 7550, 11 rue de l'Universit\'e, 67000 Strasbourg, France}

\email{frederic.marin@astro.unistra.fr}

\and
\author{Robert Antonucci}
\affil{Department of Physics, University of California, Santa Barbara, CA 93106-9530, USA}

\begin{abstract}
It is believed that, in radio-loud active galactic nuclei (AGN), the core radio flux density can be normalized 
to the flux density of the extended lobe emission to infer the orientation of a radio source. 
However very little is known about the reliability and precision of this method, and we are unaware of 
any robust conversion recipe to infer the inclination from the core dominance. Investigating whether or not the 
radio core dominance parameter R separates the quasars from the radio-galaxies in the $z \ge$~1 3CRR catalog, we 
found excellent agreement of R with optical type, infrared flux ratios and optical polarization. This indicates that 
probably both R and optical classification are very good orientation indicators, and the unified model is strongly 
predictive for these objects. The relative number densities indicate half-opening angles close to 60$^\circ$, as 
expected from large surveys. The separations of optical types according to radio core dominance as well as NIR/MIR 
ratios, which are essentially perfect, means that there can be only a small dispersion of torus half-opening angles. 
Also, even though torus dust is thought to be clumpy, there is an almost zero probability to see a type-1 source at 
high inclination. Finally, using only the Copernican Principle, i,e, the assumption that solid angle is filled 
uniformly with source axis orientations, we estimated a semi-empirical relation between core dominance and AGN 
inclination. This makes it possible to use R to infer the inclination of a source to an accuracy of $\sim$ 10 degrees 
or less, at least for this type of object. 
\end{abstract}


%%%%%%%%%%%%%%%%%%%%%%%%%%%%%%%%%%%%%%%%%%%%%%%%%%%%%%%%%%%%%%%%%%%%%%%
\section{Introduction}
\label{Intro}

The nuclear inclination of active galactic nuclei (AGN) is a fundamental parameter that still remains tricky to 
directly measure. In radio-quiet objects, proxies for measuring the orientation of AGN usually rely on degenerate 
models and fitting procedures that sometimes give inconsistent results \citep{Marin2014,Marin2016}. However, in
the case of radio-loud AGN, in which there is beamed emission from pc-scale jets and unbeamed emission from diffuse
radio lobes, we have an attractive method since the ratio of the former to the latter (the radio core dominance R) 
should be statistically a function of inclination with respect to the line of sight \citep{Orr1982}. Operationally, 
it works well because on arcsec-scale maps, double radio sources generally show an unresolved and fairly isolated 
point source, usually with a flat synchrotron spectrum implying extreme compactness (pc scales). Furthermore selection 
by the nearly-isotropic lobe emission assures that a sample has a random orientation distribution, filling all solid 
angles as seen from the nucleus. This core dominance parameter is usually estimated in the radio because, in a general 
way, the jet and lobe radiation are linked physically. But (for quasars only), the continuum optical flux density 
\citep{Wills1995} has also been used to normalize the beamed radio core flux density, instead of the flux density of 
the extended radio lobes. This core dominance parameter is called R$_{\rm V}$. Its usefulness has been confirmed by 
\citet{Barthel2000}. Narrow line luminosities have also been tried \citep{Rawlings1991}.

Very little is known about the reliability and precision of R as a statistical indicator of inclination. One way to 
test this is to see to what extent it separates radio galaxies (RG) from quasars, since we know that among 
the high-$z$ 3CR radio galaxies, all RG host hidden quasars, and that they generally have smaller inclinations. We don't know a 
priori whether or not a lot of noise will be added by a distribution of torus half-opening angles, "holes" in the torus, or 
dispersion in the radio core emission polar diagrams. But if core dominance correlates very well with object type, 
then the most natural conclusion is that the sources are roughly generic in nature, core dominance indicates inclination 
well, and there are very few quasars at high-inclinations. This is in fact what we will show\footnote{A clean 
separation of optical types by core dominance was reported by \citet{Wilkes2013} in a sample similar to ours.}. The 
unified model in its simplest form, the "Straw Person Model" of \citet{Antonucci1993}, seems to be correct to first order 
in this parameter space, as predicted by \citet{Barthel1989}. That is, the model shows very strong predictive power. The 
same conclusion follows from a detailed study of the X-ray properties of the sample by \citet{Wilkes2013}. 

High core dominance is known to be associated with low inclination and a flat IR SED; it's also known that
radio galaxies have lower core dominance than quasars; we will show that limited optical polarization is also 
consistent with this idea when available. The purpose of this paper is to note and exploit the fact that there is essentially 
zero overlap in R$_{\rm 5GHz}$ between RGs and quasars, using that fact and the special sample properties to derive
a quantitative formula for converting core dominance to inclination, estimating that it is good to plus or minus 10 degrees.
 
To obtain intelligible results, it is fundamental to select a sample were all RG have hidden quasars, and to select by an 
approximately isotropic luminosity (the radio lobes luminosity here). The high redshift (i.e. $z \ge$~1) radio-loud AGN 
from the complete flux-limited, 178~MHz selected, ``3CRR'' catalog of \citet{Laing1983} meet this requirement very well. We 
know this for several reasons. First, most these objects have been observed polarimetrically and all show substantial optical 
polarization, oriented roughly perpendicular to the radio axes \citep{Tadhunter2005}. They also show broad emission lines in 
polarized flux when that information is available\footnote{The case of 3C~368 is controversial; \citet{Dey1999} finds low 
aperture polarization, but a convincing image of centro-symmetric polarization is shown by \citet{Scarrott1990}.}. Also, all 
of these radio galaxies for which data is available show strong mid-infrared reprocessing bumps \citep{Honig2011} and high 
ionization emission lines \citep{Leipski2010}. It is important to note that this sample comprises many of the most radio-luminous 
objects in the universe, and at lower radio luminosities many radio galaxies lack energetically significant hidden quasars
\citep{Ogle2006}; the evidence for that statement from all wavebands is reviewed in detail in \citet{Antonucci2012}.

%%%%%%%%%%%%%%%%%%%%%%%%%%%%%%%%%%%%%%%%%%%%%%%%%%%%%%%%%%%%%%%%%%%%%%%
\section{Radio core dominance as a function of ...}
\label{Main}

In the following subsections, we compare radio R to the quasar/RG optical classification and also to infrared colors and 
optical polarization. Doing so, we aim to robustly constrain the reliability and precision of the radio core dominance 
parameter to build our own semi-empirical relation linking the nuclear inclination and R. 

We extracted the 3CRR object types and redshifts $z$ from the on-line catalog\footnote{http://3crr.extragalactic.info/} 
and references therein. The radio core dominance was estimated between the 178~MHz flux densities corrected to the flux 
scale used by \citet{Roger1973} and the flux density of the core at 5~GHz (where it's better isolated) on 
arcsecond scales. Additional R$_{\rm 5GHz}$ are extracted from \citet{Zirbel1995}, \citet{Fanti2002}, \citet{Fan2011}, 
and \citet{Wilkes2013}. Infrared photometry was obtained with the Spitzer Space Telescope and reported by \citet{Haas2008}. 
The optical polarization data are summarized in \citet{Tadhunter2005}. Our final, 100\% completeness sample is listed in 
Tab.~\ref{Table:Data}.

\begin{table*}
  \centering
  {
   \begin{tabular}{|c|c|c|c|c|c|c|c|c|c|}
   \hline
      {\bf 3CR}	& {\bf Type}	& {\bf $z$} 	& {\bf R$_{\rm 5GHz}$}	& {\bf $i$ ($^\circ$)} 	& {\bf $P$ (\%)} 	& {\bf $PPA$ ($^\circ$)} 	& {\bf PA ($^\circ$)}	& {\bf F$_{\rm 3.6}$ ($\mu$Jy)} 	& {\bf F$_{\rm 24}$ ($\mu$Jy)} \\
   \hline
2&	Q&	1.037&	1.905&	36 $\pm$ 10&	--&	--&	16&	283	$\pm$	42&	2970	$\pm$	446	\\				
9&	Q&	2.012&	0.25&	52 $\pm$ 10&	1.14	$\pm$	0.52&	137	$\pm$	13&	140&	884	$\pm$	133&	3470	$\pm$	520	\\
13&	RG&	1.351&	0.0007&	83 $\pm$ 10&	7	$\pm$	2&	60	$\pm$	20&	145.1&	133	$\pm$	20&	2060	$\pm$	309	\\
14&	Q&	1.469&	0.94&	42 $\pm$ 10&	7	$\pm$	2&	63	$\pm$	0&	145&	1040	$\pm$	156&	10300	$\pm$	1545	\\
43&	Q&	1.470&	$<$4.84&	$>$26&	--&	--&	155&	193	$\pm$	29&	1610	$\pm$	242	\\				
65&	RG&	1.176&	0.0005&	87 $\pm$ 10&	--&	--&	104&	202	$\pm$	30&	1700	$\pm$	255	\\				
68.1&	Q&	1.238&	0.079&	58 $\pm$ 10&	7.54	$\pm$	1.31&	52	$\pm$	5&	173&	967	$\pm$	145&	7760	$\pm$	1164	\\
68.2&	RG&	1.575&	0.01&	66 $\pm$ 10&	--&	--&	153&	105	$\pm$	16&	1170	$\pm$	176	\\				
181&	Q&	1.382&	0.38&	49 $\pm$ 10&	--&	--&	118&	348	$\pm$	52&	4260	$\pm$	639	\\				
186&	Q&	1.063&	0.97&	42 $\pm$ 10&	1.65	$\pm$	0.69&	141	$\pm$	12.96&	140&	791	$\pm$	119&	6660	$\pm$	999	\\
190&	Q&	1.197&	4.45&	27 $\pm$ 10&	--&	--&	30&	739	$\pm$	111&	6690	$\pm$	1004	\\				
191&	Q&	1.952&	2.96&	31 $\pm$ 10&	--&	--&	150&	333	$\pm$	50&	3810	$\pm$	572	\\				
204&	Q&	1.112&	2.36&	34 $\pm$ 10&	--&	--&	93&	917	$\pm$	138&	7360	$\pm$	1104	\\				
205&	Q&	1.534&	1.45&	38 $\pm$ 10&	--&	--&	14&	1460	$\pm$	219&	12800	$\pm$	1920	\\				
208&	Q&	1.109&	2.79&	32 $\pm$ 10&	1.05	$\pm$	0.5&	106	$\pm$	14&	265&	660	$\pm$	99&	5870	$\pm$	881	\\
212&	Q&	1.049&	9.1&	20 $\pm$ 10&	5.31	$\pm$	2.12&	30	$\pm$	11&	136&	925	$\pm$	139&	10800	$\pm$	1620	\\
239&	RG&	1.781&	0.0004&	89 $\pm$ 10&	--&	--&	75&	96	$\pm$	14&	1450	$\pm$	218	\\				
241&	RG&	1.617&	0.003&	72 $\pm$ 10&	--&	--&	82&	92	$\pm$	14&	591	$\pm$	89	\\				
245&	Q&	1.029&	58&	4$^{+10}_{-4}$&	0.41	$\pm$	0.58&	172	$\pm$	41&	100&	1420	$\pm$	213&	20400	$\pm$	3060	\\
252&	RG&	1.105&	0.0019&	75 $\pm$ 10&	--&	--&	105&	225	$\pm$	34&	7000	$\pm$	1050	\\				
266&	RG&	1.272&	0.0016&	76 $\pm$ 10&	--&	--&	179&	68	$\pm$	10&	980	$\pm$	147	\\				
267&	RG&	1.144&	0.0046&	70 $\pm$ 10&	--&	--&	73&	153	$\pm$	23&	3730	$\pm$	560	\\				
268.4&	Q&	1.402&	4.46&	27 $\pm$ 10&	--&	--&	160.6&	1060	$\pm$	159&	11600	$\pm$	1740	\\				
270.1&	Q&	1.519&	12.8&	16 $\pm$ 10&	--&	--&	160&	606	$\pm$	91&	5470	$\pm$	821	\\				
287&	Q&	1.055&	--&	--&	0.61	$\pm$	0.66&	119	$\pm$	31&	200&	613	$\pm$	92&	5820	$\pm$	873	\\
294&	RG&	1.786&	0.047&	60 $\pm$ 10&	--&	--&	31&	$<$93	$\pm$	0&	348	$\pm$	52	\\				
298&	Q&	1.436&	0.103&	57 $\pm$ 10&	--&	--&	78&	1600	$\pm$	240&	12600	$\pm$	1890	\\				
318&	Q&	1.574&	0.079&	58 $\pm$ 10&	--&	--&	45&	343	$\pm$	51&	3400	$\pm$	510	\\				
322&	RG&	1.681&	0.027&	63 $\pm$ 10&	--&	--&	0&	128	$\pm$	19&	804	$\pm$	121	\\				
324&	RG&	1.206&	0.0006&	85 $\pm$ 10&	18	$\pm$	1.6&	16	$\pm$	5&	71&	165	$\pm$	25&	2820	$\pm$	423	\\
356&	RG&	1.079&	0.002&	75 $\pm$ 10&	3.375	$\pm$	1.375&	30	$\pm$	20&	144&	108	$\pm$	16&	4060	$\pm$	609	\\
368&	RG&	1.132&	0.003&	72 $\pm$ 10&	2.5	$\pm$	1.2&	74	$\pm$	15&	17&	126	$\pm$	19&	3250	$\pm$	488	\\
432&	Q&	1.805&	0.62&	46 $\pm$ 10&	2.03	$\pm$	0.95&	137	$\pm$	13&	135&	420	$\pm$	63&	3940	$\pm$	591	\\
437&	RG&	1.480&	0.008&	67 $\pm$ 10&	--&	--&	162&	82	$\pm$	12&	941	$\pm$	141	\\				
454.3&	Q&	1.757&	$<$16&	$>$14&	--&	--&	--&	339	$\pm$	51&	4150	$\pm$	623	\\				
469.1&	RG&	1.336&	0.003&	72 $\pm$ 10&	--&	--&	171.1&	160	$\pm$	24&	1970	$\pm$	296	\\				
470&	RG&	1.653&	0.032&	62 $\pm$ 10&	--&	--&	37.9&	50	$\pm$	7&	2650	$\pm$	398	\\				
4C13.66&	RG&	1.450&	$<$0.16&	$>$55&	--&	--&	--&	24	$\pm$	4&	276	$\pm$	41	\\				
4C16.49&	Q&	1.296&	1.4&	39 $\pm$ 10&	--&	--&	--&	329	$\pm$	49&	1830	$\pm$	275	\\				
   \hline
   \end{tabular}
  }
  \caption{3CRR objects, types, redshifts and radio core dominance parameter R$_{\rm 5GHz}$, 
	   inferred inclination $i$, polarization degree $P$ and optical polarization position 
	   angle $PPA$, radio position angle PA (with typical error $\sim$ 10$^\circ$), 
	   and 3.6 and 24~$\mu$m fluxes from $Spitzer$. References are given in the text.
	   Incliantions $i$ are computed using the semi-empirical relation found in Sect.~\ref{Main:Relation}.}
  \label{Table:Data}
\end{table*}

\subsection{... optical classification}
\label{Main:Histogram}

\begin{figure}
   \centering
      \includegraphics[trim = 0mm 3mm 0mm 0mm, clip, width=8.7cm]{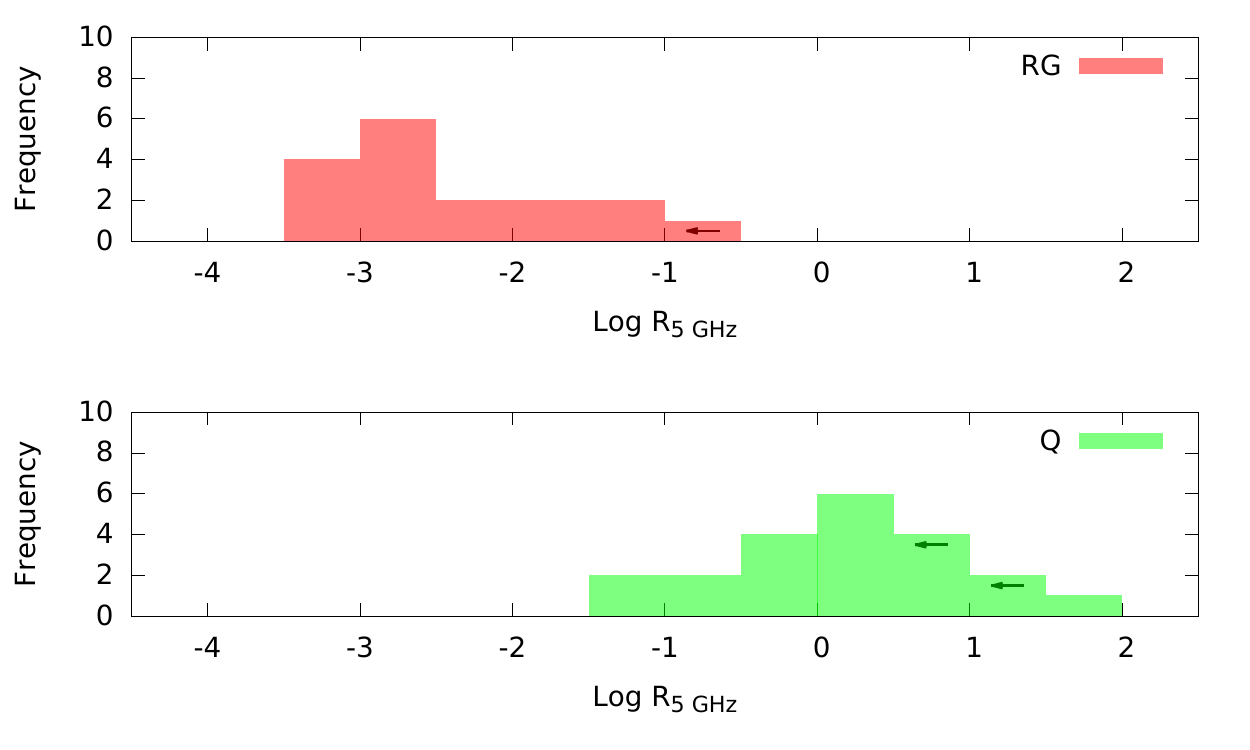}
      \caption{Number distribution of the core dominance parameter
	       for radio-galaxies (top) and quasars (bottom). 
	       Radio galaxies are shown in red, quasars 
	       in green.}
     \label{Fig:Histogram}
\end{figure}

We plot the number distribution of the core dominance parameter for radio-galaxies and quasars in 
Fig.~\ref{Fig:Histogram}. The top histogram shows the distribution of radio-galaxies (in red) and the 
bottom figure quasars (in green). The separation between the two radio-loud AGN classes is almost perfect, 
with only a few objects possibly overlapping between 0.045 $<$ R$_{\rm 5GHz}$ $<$ 0.079: two quasars 
(3C~68.1, 3C~318) and two radio-galaxies (3C~294, 4C~13.66). 

3C~68.1 is a quasar with a very low core dominance (R$_{\rm 5GHz}$ = 0.079). However, \citet{Boksenberg1976} 
have shown that its spectral index is about 6 for the optical continuum, indicating heavy absorption. 3C~68.1 
is also highly polarized, with a continuum polarization as high as 10\% in the blue band (1900~{\AA} rest frame), 
see \citet{Brotherton1998}. This indicates that 3C~68.1 is a borderline object, and this is consistent with
the core dominance being on the boundary between the two types. 3C~318\footnote{Note that 3C~318 is a compact 
steep-spectrum source showing ``a weak core with a one-sided jet to the north-east and an extended lobe to the 
south-west'' (MERLIN images, \citealt{Ludke1998}) that, for a while, was thought to be embedded in
an anomalously dense environment \citep{Fanti1995}.} has an intermediate spectral type, being classified as a galaxy by
\citet{Gelderman1994}. Broad H$\alpha$ (but not H$\beta$) is detected by \citet{Willott2000}.
This quasar shows very strong far-infrared (FIR) fluxes (148 $\pm$ 24 mJy at 60~$\mu$m) such 
as reported by \citet{Hes1995}, with FIR fluxes likely to be dominated by a non-thermal beamed component 
\citep{Hoekstra1997}. Recent observations using the Herschel Space Observatory revealed that most of the 
infrared emission from 3C~318 originates from a pair of interacting galaxies, close in projection 
\citep{Podigachoski2016}. Finally, note that the one radio galaxy plotted with R$_{\rm 5GHz}$ $>$ 0.1 is an 
upper limit. 

The overlap of quasars and radio-galaxies at 0.045 $<$ R$_{\rm 5GHz}$ $<$ 0.079 is thus naturally explained 
by one misclassified source and one object whose infrared flux is polluted by a pair of interacting 
galaxies. Accounting for these peculiarities, the resulting association between radio R and optical type is 
perfect, perhaps fortuitously so. We can draw two conclusions. First, the nature of these powerful radio-loud 
AGN is quite generic, i.e. the half-opening angle of the circumnuclear dusty region must be quite similar between 
radio-galaxies and quasars. Second, all radio-galaxies (type-2 AGN) are seen at higher inclinations than the 
quasars (type-1 AGN); there are no high-inclination quasars seen through "holes" in the torus. These conclusions 
are in perfect agreement with the unified model of AGN such as postulated by \citet{Barthel1989}, \citet{Hough1999,Hough2002}, 
and \citet{Antonucci1993}.

\subsection{... infrared fluxes}
\label{Main:IR}

Measurements from $IRAS$, $ISO$, $Spitzer$ and other infrared observatories showed that quasars have strong 
near-infrared (NIR) and MIR continua, bright silicate features and an emission bump at 2 --5~$\mu$m (see, 
e.g., \citealt{Elvis1994,Wilkes1999,Leipski2010}). Their spectral energy distribution (SED) is quite generic, 
with much lower dispersion in their fluxes and IR colour compared with radio galaxies SED \citep{Honig2011}. 
Additionally, the 2 --5~$\mu$m radio galaxy extinctions derived by \citet{Leipski2010} are higher than the 
(wavelength-normalized) values derived in the MIR, indicating that the component responsible for the NIR bump 
suffers more extinction than the MIR emission. This is consistent with the torus geometry. Hence, looking at 
the NIR and MIR fluxes it is possible to try to corroborate the association we found between R$_{\rm 5GHz}$ 
and optical type \citep{Podigachoski2015}. By selecting our sample purely on unbeamed radio emission in the 
$z \ge$~1 group, we are able to study the IR without biases. In the following, the Spitzer 3.6~$\mu$m and 
24~$\mu$m observed-frame fluxes are taken from \citet{Haas2008}, also listed in \citet{Drouart2012},
and from the NASA/IPAC Extragalactic Database\footnote{http://ned.ipac.caltech.edu/}.

\begin{figure}
   \centering
      \includegraphics[trim = 0mm 0mm 0mm 0mm, clip, width=8.7cm]{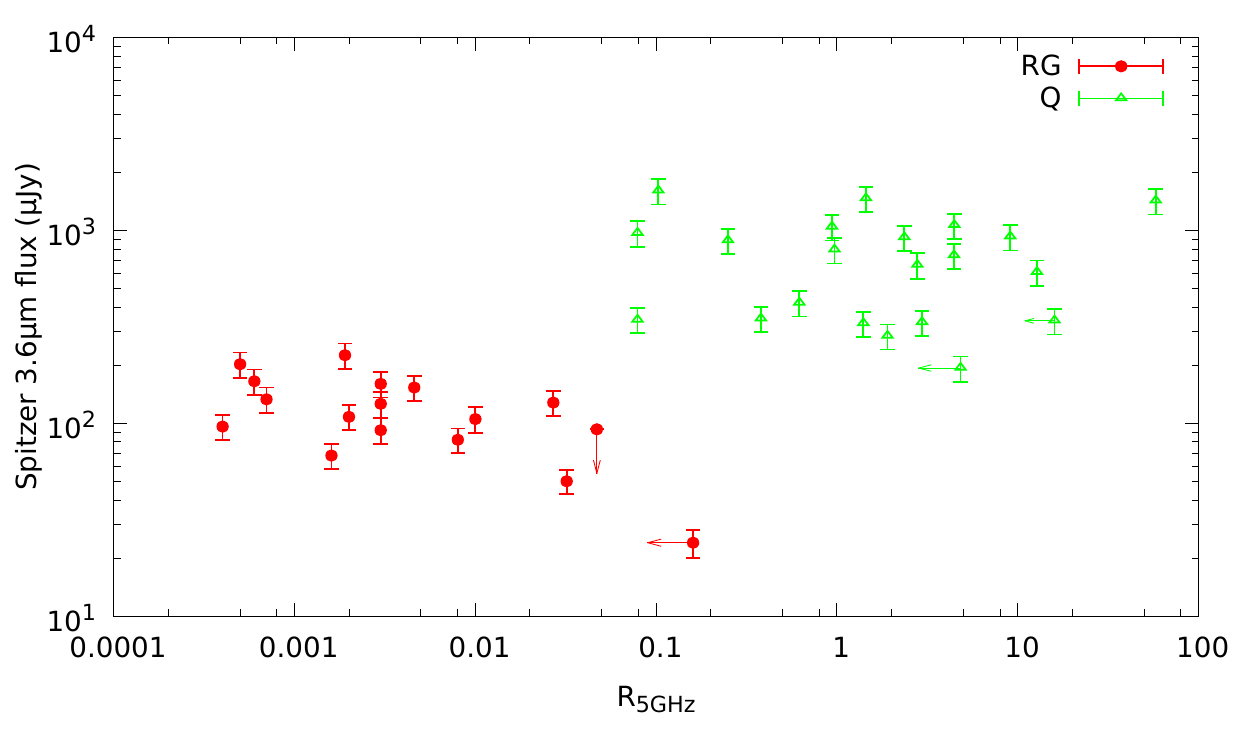}
      \includegraphics[trim = 0mm 0mm 0mm 0mm, clip, width=8.7cm]{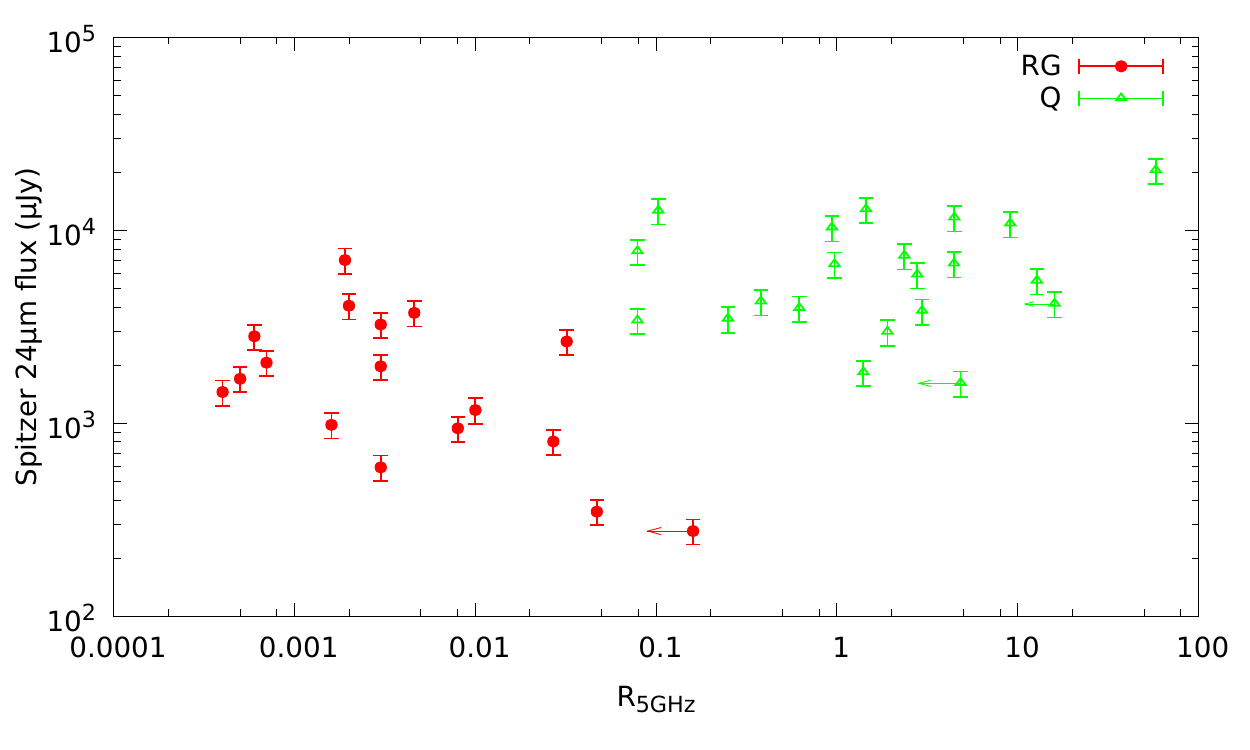}
      \caption{Core dominance versus infrared fluxes (top: 3.6~$\mu$m; 
	       bottom: 24~$\mu$m), inspired by what was presented in
	       \citet{Haas2008}. Radio galaxies are shown with red circles, 
	       quasars with green triangles.}
     \label{Fig:IR_fluxes}
\end{figure}

\begin{figure}
   \centering
      \includegraphics[trim = 0mm 0mm 0mm 0mm, clip, width=8.7cm]{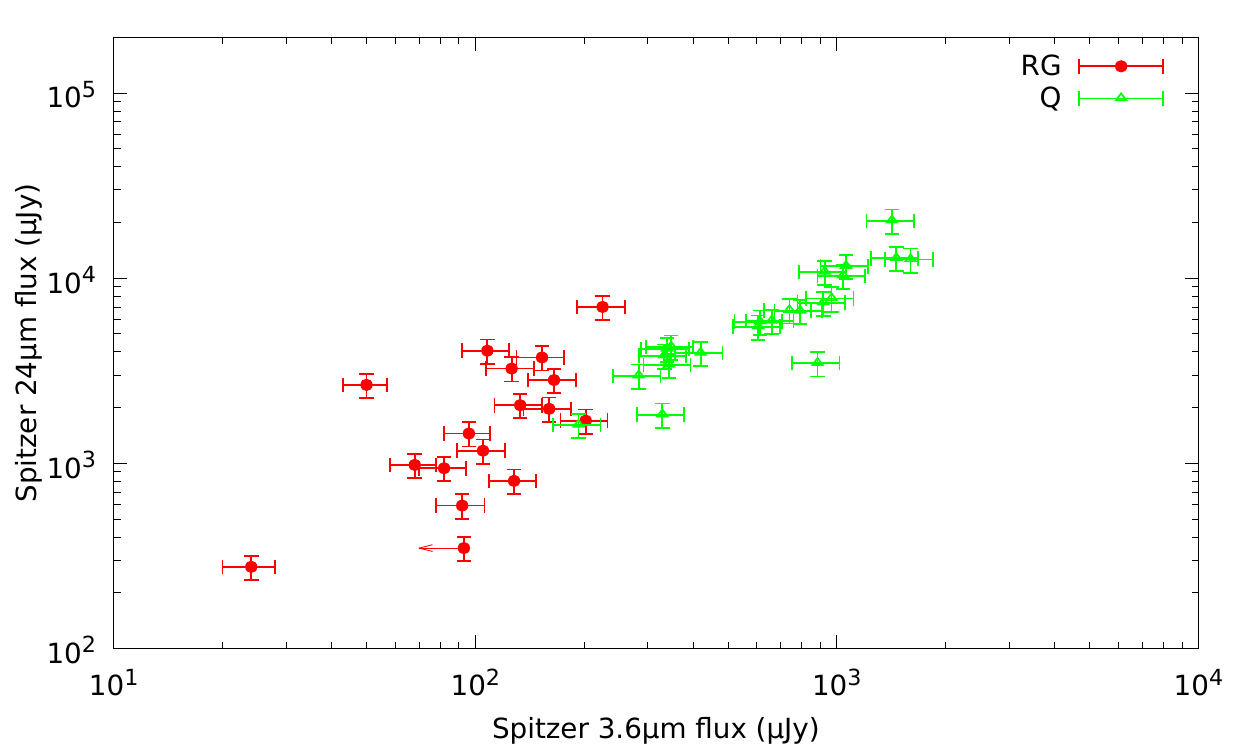}
      \caption{Plot of the near (3.6~$\mu$m) and mid (24~$\mu$m) 
	       infrared fluxes. The color code is the same as in 
	       Fig.~\ref{Fig:IR_fluxes}.}
     \label{Fig:IR_ratio}
\end{figure}

Fig.~\ref{Fig:IR_fluxes} shows the radio core dominance parameter versus NIR (top) and MIR (bottom) fluxes.
We see that the 3.6~$\mu$m RG fluxes are, on average, a factor 6.3 lower than the quasar fluxes (median: 6.8), 
which is in good agreement with the quasar/radio-galaxy flux ratios (2.5 -- 7.5) found by \citet{Honig2011}
at 3.5~$\mu$m. The same conclusions apply for the 24~$\mu$m fluxes against R$_{\rm 5GHz}$, where the MIR emission 
is found to be more isotropic. Due to the dispersion of data points, the flux ratio is more complicated to 
estimate but the averaged value is 3.3 (median: 3.4). 

The flux ratio between NIR and MIR emission is shown in Fig.~\ref{Fig:IR_ratio}. The radio-galaxies are below 
the quasars on average, by a factor $\sim$~1.35. There is a fairly low dispersion among the quasars as 
their IR SEDs are generic; in the case of radio-galaxies the distribution is less tight as RG show less uniformity 
than quasars \citep{Honig2011}. For a a given 24~$\mu$m flux, the 3.6~$\mu$m flux is much lower for the radio galaxies, 
as expected from the obscuration paradigm. Finally, it seems that we are also detecting an inclination-dependence 
of the NIR/MIR flux ratio within each class in the expected sense. 

Our conclusions that the $z \ge$ 1 3CRR radio galaxies have several times weaker NIR and MIR emission compared 
with the matched quasars, and more varied SEDs, are not new. They were stated explicitly by \citet{Haas2008}, 
\citet{Podigachoski2015}, and references therein; \citet{Honig2011} shows quotient spectra for radio galaxies 
over quasars. For lower redshift 3CR objects, \citet{Ogle2006} reported these effects, with references to much 
earlier work \footnote{In the pioneering study, \citet{Heckman1994} reported on IRAS FIR composite fluxes 
for low-redshift 3CR objects. They found that narrow line objects are on average strong emitters, but a few 
times weaker than broad line objects.  Like \citet{Barthel1989}, these authors realized that some of the
radio galaxies were poor candidates for hidden quasars based on their weak, low-ionization narrow line emission.
The difference they found was also affected by some far-IR nonthermal contamination in a few of the quasars, 
and as we now know, anisotropy due to optical depth effects.}. The novel aspect here is again to exploit
the near perfect separation of the classes by IR colors to constrain deviations from the simplest unified model.

\subsection{... and optical polarization}
\label{Main:Polarization}

The infrared fluxes have shown excellent agreement with the expectation from the unified model and strongly
corroborate our finding that there is an almost perfect separation between the two optical classes of radio-loud 
AGN based on R$_{\rm 5GHz}$. Nevertheless, we decided to do a third check using near-ultraviolet, optical
and near-IR continuum polarization measurements from the literature. The unified model again makes strong predictions 
about the degree and orientation of AGN polarization and we expect a clear dichotomy between type-1 and type-2 AGN.  
Our sample is too small to demonstrate the predicted relationships by itself, but we can check for consistency. 

\begin{figure}
   \centering
      \includegraphics[trim = 0mm 3mm 0mm 0mm, clip, width=8.7cm]{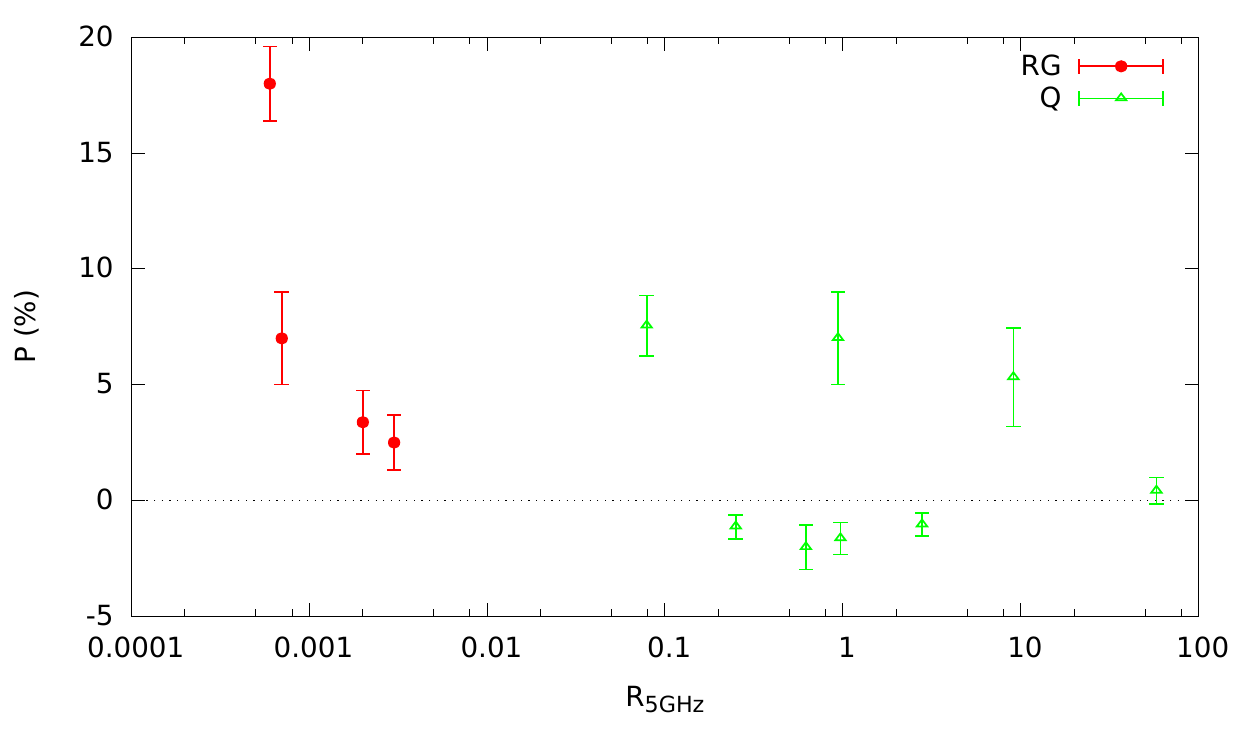}
      \includegraphics[trim = 0mm 3mm 0mm 0mm, clip, width=8.7cm]{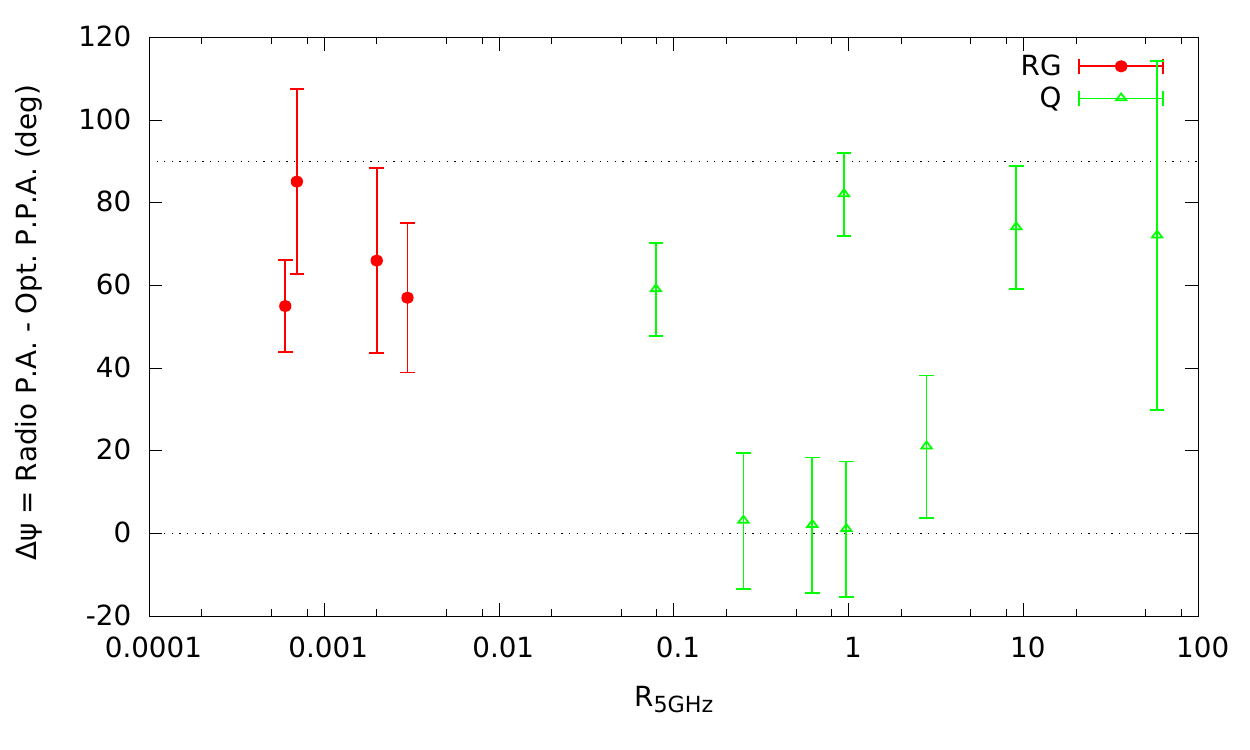}
      \caption{Optical continuum polarization versus core 
	       dominance. Top: linear polarization degree $P$; 
	       bottom: difference ($\Delta\psi$) between the 
	       radio position angle PA and the optical polarization 
	       position angle $PPA$. We use the following 
	       convention: $P$ is plotted as negative when 
	       the $\vec E$-vector of the continuum radiation 
	       aligns with the projected axis of the (small-scale) 
	       radio structures ($PPA$ = 90$^\circ$), and $P$ 
	       is positive for perpendicular polarization 
	       ($PPA$ = 0$^\circ$). The horizontal lines show 
	       the minimum and maximum realizable value for 
	       the position angle difference. The color 
	       code is the same as in Fig.~\ref{Fig:IR_fluxes}.}
     \label{Fig:Polarization}
\end{figure}

Tab.~\ref{Table:Data} lists the only 13 polarimetric data reported for the 3CRR objects in our sample. These AGN 
have been observed in U, B, V and/or R polarization filters, and were corrected for the small polarizations of the 
instrument and the Galactic interstellar medium. The polarization degree $P$ and the difference $\Delta\psi$ between 
the radio position angle PA and the optical polarization position angle $PPA$ are presented in Fig.~\ref{Fig:Polarization} 
top and bottom, respectively. $P$ seems similar between RG and quasars, but radio galaxies are generally substantially 
diluted by starlight, while that of quasars is not \citep{Cimatti1997}, which is part of the reason why the polarization 
degree of the two groups seems similar. However, there is a net difference between type-1 and type-2 AGN: RG only show a 
polarization position angle approximately perpendicular to the projected axis of the radio structures ($\Delta\psi$ = 90$^\circ$) while 
quasars are characterized by a blend of perpendicular and parallel polarization angles. The exact same dichotomy appears 
in radio-quiet AGN \citep{Antonucci1993} and radio-loud AGN at lower redshifts \citep{Antonucci1984}, where type-2s have 
perpendicular $PPA$ only. The presence of perpendicular $PPA$ indicates that photons dominantly experience scattering in 
polar directions, while a parallel $PPA$ is the signature of equatorial scattering. Polar scattering dominated quasars 
tend to have higher degrees of polarization (up to a couple of percents) than equatorial scattering quasars, as already 
observed for their radio-quiet analogues \citep{Smith2002}. A large part of the higher polarization in well-studied 
type-2 AGN results from the fact that the direct light from the quasar is blocked; otherwise it would dilute the observed 
polarization, often very strongly. However, the relatively high polarization of the polar-scattered quasars indicates that
our inclination-dependent view of the scattering medium also plays a role. Finally, we note that the incidence of high 
parallel polarization in the quasars is much higher than in other well-studied samples \citep{Stockman1979}, 
although it's not clear why this should be the case.

\section{A semi-empirical function of core dominance as a function of inclination}
\label{Main:Relation}

Since we are very confident that core dominance is a good qualitative inclination indicator, and that the sources 
are distributed uniformly in solid angle (Copernican Principle), we can make a quantitative semi-empirical function 
for the average inclination $i$ as a function of core dominance. It should be of interest to both observers and 
theorists \citep{Blandford1979}. To do so, we equally distributed the sources in solid angle and numerically fitted 
the relation using a polynomial regression. We obtained the following semi-empirical relation between core dominance 
and AGN inclination:

\[
    log({\rm R}_{\rm 5GHz}) = {\rm a} + {\rm b}i + {\rm c}i^{2} + {\rm d}i^{3} + {\rm e}i^{4} + {\rm f}i^{5}, \label{Ri}
\]

or, if numerically inverted\footnote{Fifth degree polynomials are generally problematic to invert; the behavior 
of the inverted function $i$(R$_{\rm 5GHz}$) becomes less reliable for extremes values of $i$.} :

\[
    i = {\rm g} + {\rm h}(LR) + {\rm j}(LR)^{2} + {\rm k}(LR)^{3} + {\rm l}(LR)^{4} + {\rm m}(LR)^{5}, \label{iR}
\]

with LR = $\log({\rm R}_{\rm 5GHz})$. The first formula, essentially the polar diagram for core emission, 
should be more useful to theorists, while the second version is directly useful to observers wishing to estimate 
the inclination for a given system.

\begin{table}
  \centering
  {
   \caption{Parametrization of the fits and goodness measures}
    \begin{tabular}{lr}
    \textbf{Parameter for R(i)} & \textbf{Value} \\
    \hline
    a 	&  2.916 $\pm$ 0.124  \\
    b   & -0.225 $\pm$ 0.027  \\
    c   &  0.009 $\pm$ 0.002  \\
    d   & -1.799$\times$10$^{-4}$ $\pm$ 4.987$\times$10$^{-5}$  \\
    e   &  1.347$\times$10$^{-6}$ $\pm$ 5.941$\times$10$^{-7}$  \\    
    f   & -2.927$\times$10$^{-9}$ $\pm$ 2.557$\times$10$^{-9}$  \\    
    \hline
    \textbf{Goodness measures} & \textbf{Value} \\  
    \hline  
    $\cal{R}$$^2$\footnotemark[\textdagger]		&  0.9986 \\
    a$\cal{R}$$^2$\footnotemark[\textdaggerdbl]		&  0.9978 \\
    SE\footnotemark[$\|$] (in units of R$_{\rm 5GHz}$)	&  0.0816 \\
    \hline
    \textbf{Parameter for i(R)} & \textbf{Value} \\
    \hline
    g 	&  41.799 $\pm$ 1.190  \\
    h   & -20.002 $\pm$ 1.429  \\
    j   &  -4.603 $\pm$ 1.347  \\
    k   &   0.706 $\pm$ 0.608  \\
    l   &   0.663 $\pm$ 0.226  \\    
    m   &   0.062 $\pm$ 0.075  \\    
    \hline
    \textbf{Goodness measures} & \textbf{Value} \\  
    \hline  
    $\cal{R}$$^2$	&  0.9931 \\
    a$\cal{R}$$^2$	&  0.9905 \\
    SE (in units of $i$)&  2.7244 \\   
    \hline
    \end{tabular}	
    \footnotetext[\textdagger]{$\cal{R}$$^2$ is the ratio of variation that is 
		  explained by the curve-fitting model to the total 
		  variation in the model.}
    \footnotetext[\textdaggerdbl]{a$\cal{R}$$^2$ is the $\cal{R}$$^2$ value
		  adjusted downward to compensate for over fitting. If a model 
		  has too many predictors and higher order polynomials, it begins 
		  to model the random noise in the data. This condition produces 
		  misleadingly high $\cal{R}$$^2$ values and a lessened ability 
		  to make predictions.}   
    \footnotetext[*]{SE is the standard error (i.e. the 
		  root-mean-square of the residuals) of the formula 
		  relative to the observed distribution, and is not 
		  the absolute uncertainty of i or R derived from 
		  the formula, which is discussed in the text.}
  }
  \label{Tab:Fit}
\end{table}

\begin{figure}
   \centering
      \includegraphics[trim = 0mm 3mm 0mm 0mm, clip, width=8.7cm]{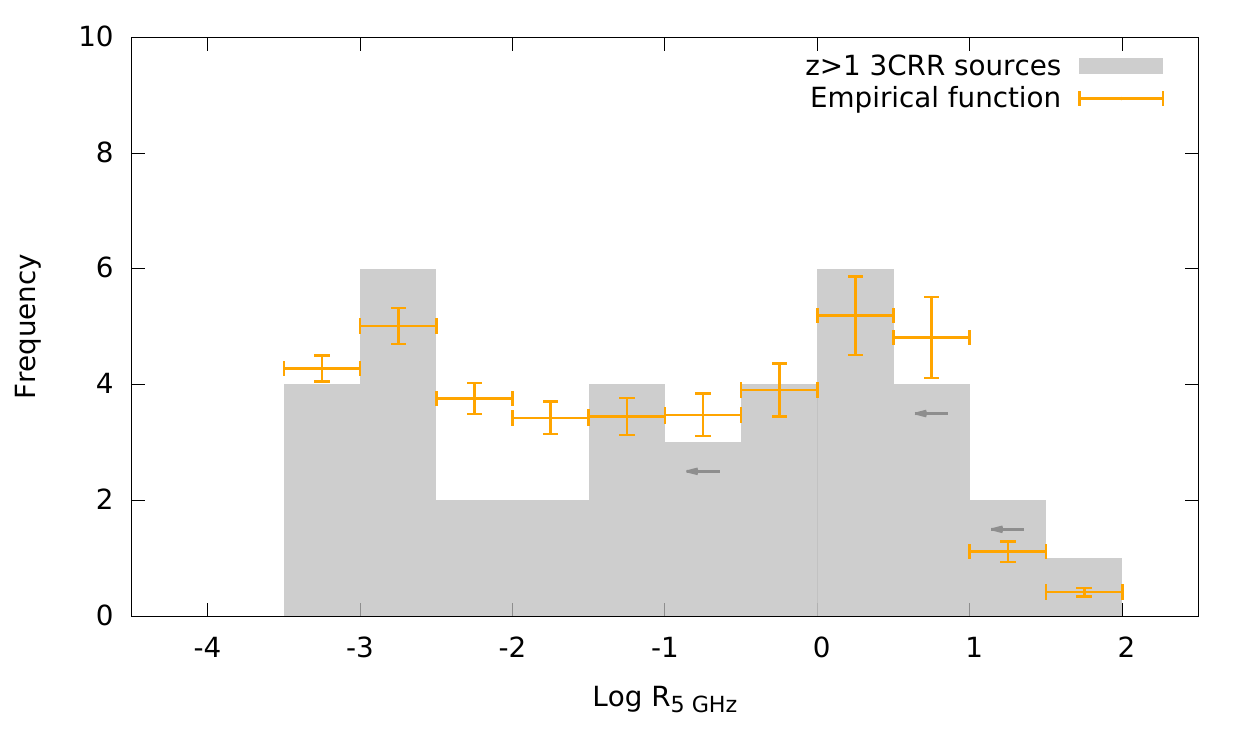}
      \includegraphics[trim = 0mm 0mm 0mm 2mm, clip, width=8.7cm]{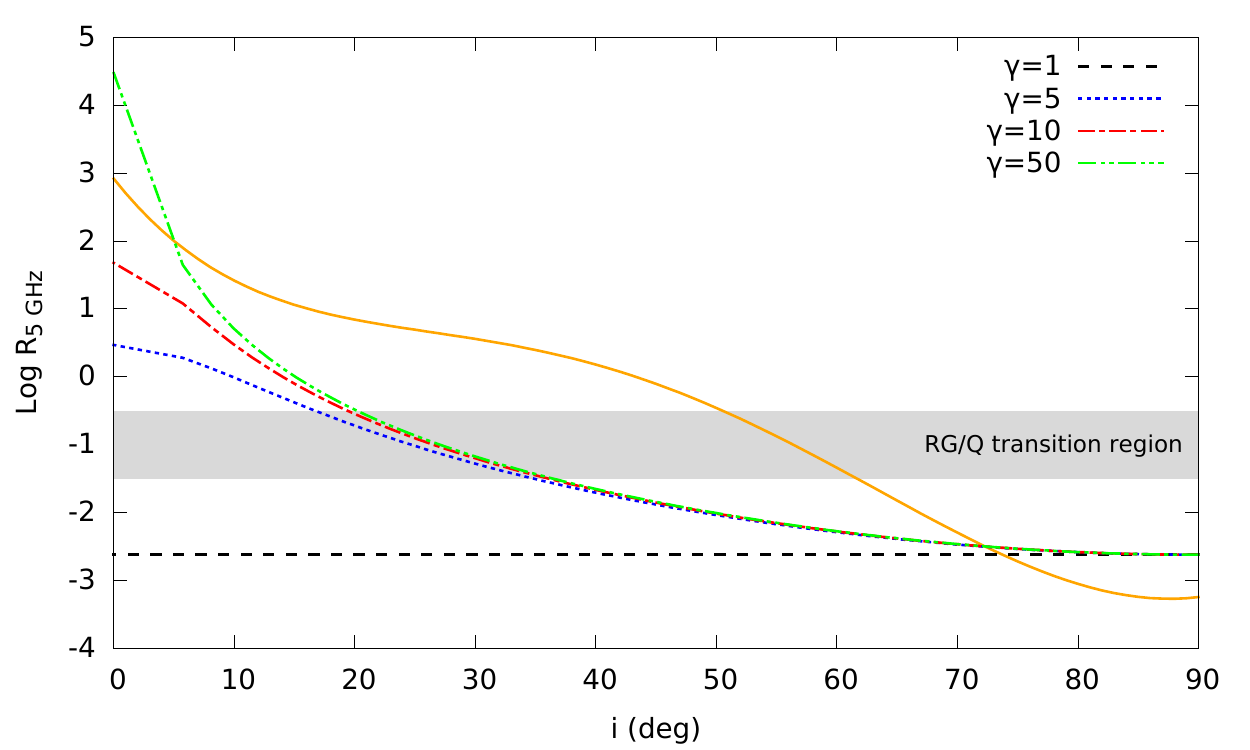}
      \caption{Comparing our semi-empirical function (solid orange line) with
	       the 3CRR sample (top figure) and the relativistic beaming model 
	       (bottom) presented in \citet{Orr1982} for different $\gamma$ factors.}
     \label{Fig:Theory}
\end{figure}

The parameters ``a'' and ``g'' are constants that represent the equivalent emission of an isotropic component. 
Values for $a$, $b$, $c$, $d$, $e$, $f$, $g$, $h$, $j$, $k$, $l$, and $m$,  are given in Tab.~\ref{Tab:Fit}, together
with the goodness measures of the fitting function to the distribution of the sources in solid angle. The typical 
uncertainty in $i$ can be no more than about $\pm$ 10$^\circ$; otherwise simulations show that the RG and quasars would 
overlap substantially. The number distribution resulting from this analytical formula is shown in 
Fig.~\ref{Fig:Theory} (top). Our new function properly reproduces the observed increase of quasar number at 
large R$_{\rm 5GHz}$ and predicts that the distribution should be almost uniform in $i$ at large inclinations. Note 
that our formula must not be used in highly core-dominant objects, since the distribution is not sampled well at small 
inclinations.

Although we don't have many sources, we can estimate the averaged half-opening angle of the circumnuclear region by 
estimating the ratios of type-1 to type-2 objects in out complete sample, which is 0.574, corresponding to a critical 
inclination of 55$^\circ$ (60$^\circ$ if excluding the two peculiar quasars mentioned in Sect.~\ref{Main:Histogram}). 
This result agrees with the values found by previous authors, including \citet{Barthel1989},  \citet{Arshakian2005}, 
\citet{Sazonov2015} and many others, who looked at high-luminosity radio-loud AGN and estimated that the half-opening 
angle of the equatorial region is $\sim$ 45$^\circ$ -- 60$^\circ$. In the future we may enlarge our sample, or explore 
other parameter space using the same methods. 

We plotted R$_{\rm 5GHz}$ as a function of $i$ in Fig.~\ref{Fig:Theory} (bottom), together with the analytical 
formula from \citet{Orr1982}\footnote{The original formula is by \citet{Scheuer1979}. \citet{Orr1982} 
included the contribution of the receding side of the compact core. Note that the formula is based only
on special relativity and is independent on the actual classification of RG and quasars.}, where it is assumed 
that jets are co-linear (zero opening angle) and moving with a single bulk velocity, so that R$_{\rm 5GHz}$($i$) 
is just determined by special relativity. When compared with this naive model, our semi-empirical formula shows 
much stronger than predicted core fluxes at intermediate angles, so that the polar diagram is not as sharply peaked 
at $i$ = 0 as in the toy model. Taken at face value, the bulge in the polar emission diagram 
at $i \sim$ 40$^\circ$ is consistent with the popular spine-and-sheath model \citep{Sol1989}, where there is 
a slow jacket to a fast very narrow jet. The slower sheet (layer) is only modestly beamed and substantial broadening 
of the effective beaming cone is required. This indicates a range of bulk speeds among objects or within individual 
jets, a range of direction of synchrotron-emitting plasma elements, optical depth effects, or other deviations from 
the single-zone jet \citep{Blandford1979}. However, our sample is small and better statistics are needed to 
evaluate the statistical significance of the bulge and, more generally, of the R(i) formula.

%%%%%%%%%%%%%%%%%%%%%%%%%%%%%%%%%%%%%%%%%%%%%%%%%%%%%%%%%%%%%%%%%%%%%%%
\section{Summary, discussion, and conclusions}
\label{Conclusion}

We have demonstrated that the radio core dominance parameter R$_{\rm 5GHz}$ separates the radio-galaxies and 
quasars almost perfectly in the $z \ge$~1 3CRR sample. In other words, the agreement of R$_{\rm 5GHz}$ with 
optical type is not a coincidence, and probably both R$_{\rm 5GHz}$ and optical type are reliable orientation 
indicators. Since the 3CRR catalog is a complete sample, it must fill the solid angle uniformly, except for small 
number statistics, just from the Copernican Principle. Therefore we were able to derive an empirical core dominance formula, 
where R$_{\rm 5GHz}$ is a function of inclination $i$. The essentially perfect separation of the optical types by 
radio core dominance and infrared (and X-ray columns) is most simply interpreted as meaning that there is only a 
small dispersion of opening angle, and no holes in the circumnuclear region which would let us see a type-1 AGN at 
high inclination (contrary to the hypothesis of \citealt{Obied2016}). 

At lower luminosities, such predictions are less clear as the sample gets less clean. It is possible to extend the 
catalog to lower redshifts but it would be necessary to exclude nonthermal RG. Weakly-accreting radio-galaxies mostly 
radiate through kinetic energy in the form of synchrotron jets and lack highly ionized line emission and strong IR 
reprocessing bumps, which indicates that they lack energetically significant hidden quasars (\citealt{Antonucci2012} 
and references therein). It is then essential to remove them to have a correct sample.

At higher redshifts another problem may arise due to the fact that the ratio of lobe emission to jet power is sensitive 
to environment \citep{Barthel1996}. The density of the intergalactic medium (IGM) scales with redshift as (1 + $z$)$^3$ 
\citep{Macquart2013}. For a $z$ = 5 quasar, the IGM density is 27 times higher than compared to a $z$ = 1 quasar, and 
these should strongly affect the morphology, lobe flux ratios, and lobe distance.

\acknowledgements
The authors would like to thank the anonymous referee for her/his useful and constructive comments on the 
manuscript. We are grateful to Belinda Wilkes, Peter Barthel, Patrick Ogle, Pece Podigachoski, and Alan Marscher 
for their nice suggestions that helped to improve the quality of this paper.

%%%%%%%%%%%%%%%%%%%%%%%%%%%%%%%%%%%%%%%%%%%%%%%%%%%%%%%%%%%%%%%%%%%%%%%

\label{lastpage}

\end{document}